\documentclass[aps,prc,showpacs,floatfix,footinbib,twocolumn,superscriptaddress]{revtex4-1}

\usepackage{amsmath}
\usepackage[dvips]{graphics,graphicx}
\usepackage{graphicx}
\usepackage{dcolumn}
\usepackage{bm}
\usepackage{latexsym}
\usepackage{graphics}
\newcommand{\be} {\begin{equation}}
\def\ee {\end{equation}}
\def\nn {\nonumber}
\def\bea {\begin{eqnarray}}
\def\eea {\end{eqnarray}}

\def\1x{\xi^{(1)}}
\def\2x{\xi^{(2)}}
\def\3x{\xi^{(3)}}
\def\4x{\xi^{(4)}}
\def\5x{\xi^{(5)}}
\def\6x{\xi^{(6)}}
\def\7x{\xi^{(7)}}
\def\8x{\xi^{(8)}}

\begin{document}
\title{One particle distribution function and shear viscosity in magnetic field: a relaxation time approach}
\author{Payal Mohanty }
 \email{payal.mohanty@gmail.com}
\author{Ashutosh Dash}
 \email{ashutosh.dash@niser.ac.in}
\author{Victor Roy}%
 \email{victor@niser.ac.in}
\affiliation{%
National Institute of Science Education and Research,
HBNI, 752050 Odisha, India.\\
}

\begin{abstract}
We calculate the $\delta f$ correction to the one particle distribution function in 
presence of magnetic field and non-zero shear viscosity within the relaxation time approximation. 
The $\delta f$ correction is found to be electric charge dependent.
Subsequently, we also calculate one longitudinal and four transverse shear viscous coefficients
as a function of dimensionless Hall parameter $\chi_{H}$ in presence of the magnetic field.
We find that a proper linear combination of the shear viscous coefficients calculated in this work 
scales with the result obtained from Grad's moment method in \cite{Denicol:2018rbw}.
Calculation of invariant yield of $\pi^{-}$ in a simple Bjorken expansion with cylindrical 
symmetry shows no noticeable change in spectra due to the $\delta f$ correction for realistic 
values of the magnetic field and relaxation time. However, when transverse expansion is taken 
into account using a blast wave type flow field we found noticeable change in spectra and elliptic 
flow coefficients due to the $\delta f$ correction. The $\delta f$ is also found to be very sensitive 
on the magnitude of magnetic field. Hence we think it is important to take into account the $\delta f$ 
correction in more realistic numerical magnetohydrodynamics simulations.
\end{abstract}

\maketitle

\section{Introduction}
\label{sec:level1}
Ultra-intense transient electromagnetic fields are generated in the initial stages 
of high energy heavy ion collisions \cite{Bzdak:2011yy,Deng:2012pc,Roy:2015coa,Kharzeev:2007jp,Rafelski:1976,Li:2016tel,Tuchin:2013apa}.
The possibility of the existence of such an 
intense magnetic field has encouraged theoretician to study QCD under the intense field.
This has so far resulted in a number of a conjectured new phenomenon which is believed to 
exist in presence of an intense electromagnetic field. For example, the phenomena of
chiral magnetic effect~(CME), chiral magnetic wave, and change in the photon and dilepton productions 
to name a few~\cite{Kharzeev:2007jp,Tuchin:2014iua,Tuchin:2010vs,Tuchin:2010gx,Basar:2012bp,Hirono:2014oda}.
Some other theoretical developments related to the magnetic field in heavy ion collisions  includes  
magneto-vortical evolution \cite{Dash:2017rhg,Florkowski:2018ubm,McInnes:2018hid},
calculation of Wigner functions for fermions in strong magnetic fields \cite{Sheng:2017lfu},
shear viscosity in an anisotropic unitary Fermi gas\cite{Samanta:2016pic},
the shear and bulk viscosity of Quark-Gluon-Plasma~(QGP) in strong magnetic fields 
\cite{Li:2017tgi,Hattori:2017qih,Ghosh:2018cxb} etc.
On the other hand, several model studies in the last decade show that the QGP
created in high energy heavy ion collisions possesses a very small value of shear viscosity to 
entropy density ratio.
Among several of these model studies, relativistic hydrodynamics played one of the most important roles 
to extract the value of shear viscosity to entropy density ratios from the available experimental data \cite{Shen:2011zc,Romatschke,Heinz,Bozek:2012qs,Roy:2012jb,Niemi:2012ry,Heinz:2011kt,Schenke:2011bn,Romatschke:2017ejr}.
However, almost all of these hydrodynamics model studies have so far ignored 
the effect of a large magnetic field on the fluid evolution, hence, the extracted values of 
shear viscosity are probably not as precise as is usually claimed. Only in the recent year's 
people finally start investigating the effect of magnetic field on QGP 
evolution \cite{Gursoy:2014aka,Zakharov:2014dia,Pang:2016yuh,Inghirami:2016iru,Das:2017qfi,Greif2017,Roy:2015kma,Roy:2017yvg,Pu:2016ayh,Pu:2016bxy,Stewart:2017zsu,Moghaddam:2017myy}. In addition to that theoretical efforts are going on to better understand the relativistic magnetohydrodynamics from first principle calculations, for example the formulation of relativistic 
non-resistive dissipative magnetohydrodynamics from a kinetic theory approach can be found in \cite{Denicol:2018rbw},
and some application to astrophysical problems is described in \cite{Huang:2009ue}, the calculation of transport coefficients in magnetic field using Kubo formula was investigated in \cite{Huang:2011dc}.
Here  we would like to point out that almost all of the recent numerical hydrodynamic model studies 
with a non-zero magnetic field have concentrated only on the effect of the field on the fluid evolution.
The effect of the magnetic field on the freezeout distribution function and hence on the 
corresponding correction to the invariant yield has so far been neglected in all of those 
magneto hydrodynamical model studies. 
As it is well known, the freezeout distribution function is used in 
the Cooper-Frye prescription to convert the fluid elements to particles~(hadrons) during 
the kinetic freezeout in order to get the invariant yields of particle spectra.
In the present study, we use the relaxation time approximation to calculate the $\delta f$ 
correction to the one particle equilibrium distribution function $f_{0}$ and the corresponding 
shear viscosity coefficients in presence of a magnetic field. It is also known that under the influence of 
an external magnetic field locally equilibrated thermal system becomes anisotropic
and corresponding shear viscosity has five different coefficients. We would  like to reiterate that
the motivation for calculating the $\delta f$ correction arises to complete the study of
the transverse momentum spectra and the corresponding flow harmonics calculated in 
hydrodynamics simulation with non-zero magnetic field. However, in this study we shall use the
calculated $\delta f$ in a simplified Bjorken model with and without transverse flow 
to investigate the effect on invariant yield 
coming only from the $\delta f$.
Before proceeding further, we note that the local equilibrium distribution function in presence of 
the electromagnetic field is known to have the following close form \cite{Degroot:book,Hakim:Pr}
\begin{equation}
\label{eq:EqDfuncEMF}
    f_{0}^{em}(p)=\frac{1}{(2\pi)^3}exp\left(-\beta\left[(p^{\mu}+qA^{\mu})u_{\mu}-\mu\right]\right),
\end{equation}

where $q$ is the electric charge, $A^{\mu}$ is the four potential corresponding to an 
electro-magnetic field, and $\mu$ is the chemical potential. 
We also note, that $A^{\mu}$ is not uniquely defined for an arbitrary given
magnetic field (or in other word using a different gauge a new $A'^{\mu}$
can also give the same magnetic field as before) and this ambiguity in defining $A^{\mu}$
makes it difficult to use $f_{0}^{em}(p)$ in the Cooper-Frye formula 
\begin{equation}
\label{eq:CF}
    E\frac{d^3N}{d^3p}=\int f_{0}^{em}p^{\mu}d\Sigma_{\mu}.
\end{equation}
Where $d\Sigma_{\mu}$ is the differential freezeout hypersurface, and 
$p^{\mu}\rightarrow p^{\mu}+qA^{\mu}$ is the canonical momentum.

Thus we cannot use Eq.(\ref{eq:EqDfuncEMF}) in the Cooper-Frye freezeout 
formula Eq.(\ref{eq:CF}) in order to study the effect of magnetic field
on the freezeout distribution function.
Therefore we choose a different approach and calculate the correction $\delta f$ to the local 
equilibrium distribution function $f_{0}(p)$ in presence of an external magnetic field 
by considering the $\delta f$ to be small in comparison to the $f_{0}(p)$. 

The paper is organized as follows. In the next section, we discuss the formulation
where $\delta f$ is calculated from the linearised Boltzmann transport equation 
in presence of a non-zero velocity gradient and magnetic field. This section also contains some results 
on particle spectra in Bjorken expansion with and without transverse flow and we discuss the shear viscosity for Bose and Fermi 
gas in presence of the magnetic field. Finally, in section III we discuss conclusion 
and outlook. We use the natural unit where $\hbar=c=k_{B}=1$ and four vectors are
denoted by Greek indices and three vectors by Latin indices.

\section{\label{sec:level2} Formalism}

For the sake of completeness, let us start with the relativistic Boltzmann transport equation: 
\begin{equation}
\label{eq:RelBoltz}
    p^{\mu}\frac{\partial f}{\partial x^{\mu}} + m_{0}\mathcal{F}^{\mu}\frac{\partial f}{\partial p^{\mu}}=\mathcal{C}[f],
\end{equation}
where $m_{0}$ is rest mass, $p^{\mu}=\gamma (m_{0},m_{0}\vec{v})=(p^{0},\vec{p})$ is the four momentum and 
$\mathcal{F}^{\mu}=\gamma (\vec{F}\cdot \vec{v},\vec{F})=(\mathcal{F}^{0},\vec{\mathcal{F}})$ is the four force.
$\mathcal{C}[f]$ on the right hand side of Eq.(\ref{eq:RelBoltz}) is the collision integral which is given 
by (with the assumption of hypothesis of molecular chaos and binary collisions)
\begin{eqnarray}
\nonumber
\mathcal{C}[f]&=&\frac{1}{2}\int\frac{d^3p_{1}}{p_{1}}\frac{d^3p\prime}{p\prime}\frac{d^3p\prime_{1}}{p\prime_{1}}\\
& &\left[f\prime f\prime_{1} w(p\prime,p\prime_{1}|p,p_{1})-ff_{1}w(p,p_{1}|p\prime,p\prime_{1})\right],
\end{eqnarray}
where $f,f_{1},f\prime$ and $f\prime_{1}$ stands for one particle distribution 
$f(x^{i},p^{i}),f(x^i,p_{1}^{i}),f(x^{i},p\prime^{i})$ and $f(x^{i},p\prime_{1}^{i})$ respectively,\newline
$w(p\prime,p\prime_{1}|p,p_{1})$ is the transition rate. In equilibrium, the collision integral 
$\mathcal{C}[f_{0}]$ vanishes for the equilibrium distribution function 
$f_{0}(x^{i},p^{i})$. In principle one can evaluate corresponding 
collision integral and solve the Boltzmann transport equation. 
However, the general form of 
collision integral is not easy to handle and usually, it is non-trivial
to solve the transport equation analytically. 
A great deal of simplification can be made if we replace the collision 
integral by $\mathcal{C}[f]=-\frac{f-f_{0}}{\tau_{c}}$, where $\tau_{c}$ is the 
relaxation time, during which the system slightly away from equilibrium, 
relaxes back to the nearest equilibrium state. 
In other words, for systems slightly away from the equilibrium
the distribution function can be approximated as $f=f_{0}+\delta f$, 
where $\delta f$ denotes the deviation from the equilibrium 
distribution function with the additional assumption 
of $\frac{\delta f}{f_{0}}\ll 1$. The calculation of $\delta f$
and hence kinetic coefficients (viscosity) can be simplified 
by noting that they don't depend on the fluid velocity $\bf{V}$ explicitly. 
It is sufficient to consider at any point in the fluid where $\bf{V}$
is zero but has non-zero spatial derivative i.e., we consider the fluid rest frame.
In this case one can express the Boltzmann equation in relaxation time 
approximation by considering only magnetic force and shear viscosity as
\cite{Landau:book10,Ofengeim:2015qxz}
\begin{equation}
\label{eq:Boltzman2}
    \left(v_{i}p_{j}\frac{\partial V_{i}}{\partial x_{j}}-\frac{1}{3}v_{l}p^{l}{\bf \nabla}\cdot {\bf V}\right)\left(\frac{\partial f_{0}}{\partial \epsilon}\right)=-\frac{\delta f}{\tau_{c}}+q\varepsilon_{ijk}v_{j}B_{k}\frac{\partial \delta f}{\partial p_{i}},
\end{equation}
where q is the electric charge, and $\varepsilon_{ijk}$ is the totally antisymmetric Levi-civita tensor.
The last term in Eq.(\ref{eq:Boltzman2}) corresponds to the velocity dependent magnetic force on charged
particles. As mentioned earlier, here we consider the $\delta f$ correction for the non-zero velocity gradient
and magnetic field, the corresponding viscous stress tensor $\sigma^{ij}$ is proportional to the symmetric 
stress tensor of fluid velocity as 
\begin{equation}
 \label{eq:stressTens}
    \sigma^{ij}=\eta^{ijkl}V_{kl}
\end{equation}
where $\eta^{ijkl}$ is the viscosity tensor,
\begin{equation}
V^{kl}=\frac{1}{2}\left(\frac{\partial V^{k}}{\partial x^{l}}+\frac{\partial V^{l}}{\partial x^{k}}\right)    
\end{equation}
and the fluid velocity ${\bf V}$ is assumed to be non-relativistic. If we consider only shear viscosity (zero bulk viscosity)
in a magnetic field ${\bf B}$ with the unit magnetic field vector ${\bf b}=\frac{\bf B}{|B|}$ then Eq.(\ref{eq:stressTens})
can be written as 
\begin{equation}
\label{eq:etanSn}
    \sigma^{ij}=\sum_{n=0}^{4}\eta_{(n)} S^{ij}_{(n)}.
\end{equation}
The above expression is constructed in such a way that each of the second rank tensor $S^{ij}_{(n)}$ (for $n=0-4$) 
gives zero on contraction with respect to the indices $i,j$.
$\eta_{(0)},\eta_{(1)},\allowbreak\eta_{(2)}, \eta_{(3)},\eta_{(4)}$ are shear
viscosity co-efficients. The second rank symmetric trace zero tensor $S^{ij}_{(n)}$'s are constructed out of 
$\delta^{ij}$, $V^{ij}$,$b^{i}$, $b^{j}$, and $b^{ij}$, where 
$b^{ij}=\epsilon^{ijk}b_{k}$.
\begin{eqnarray}
\label{eq:trlessShr1}
S^{ij}_{(0)}&=&\left(3b^{i}b^{j}-\delta^{ij}\right)\left(b^{j}b^{k}V_{jk}-\frac{1}{3}{\bf \nabla}\cdot{\bf V}\right), \\
\nonumber
\label{eq:trlessShr2}
S^{ij}_{(1)}&=& 2V^{ij}-\delta^{ij}{\bf \nabla \cdot V} -2V^{ik}b_{k}b^{j}-2b^{i}V^{jk}b_{k}+ \\
            & & \delta^{ij}V^{kl}b_{k}b_{l}+b^{i}b^{j}{\bf \nabla \cdot V}+b^{i}b^{j}V^{kl}b_{k}b_{l}, \\
\label{eq:trlessShr3}            
S^{ij}_{(2)}&=& 2\left(V^{ik}b_{k}b^{j}+b^{i}V^{jk}b_{k}-2b^{i}b^{j}V^{kl}b_{k}b_{l}\right), \\ 
\label{eq:trlessShr4}
S^{ij}_{(3)}&=& -b^{ik}V_{k}^{j}-V^{ik}b_{k}^{j}+b^{ik}V_{kl}b^{l}b^{j}+b^{i}b^{jk}V_{kl}b^{l} , \\
\label{eq:trlessShr5}
S^{ij}_{(4)}&=& -2\left(b^{ik}V_{kl}b^{l}b^{j}+b^{i}b^{jk}V_{kl}b^{l}\right).
\end{eqnarray}
Viscous coefficient $\eta_{(0)}$ in Eq. (\ref{eq:etanSn}) is called longitudinal shear viscosity since $S^{ij}_{(0)}b_{i}b_{j}\neq 0$,
for similar reasons rest of the $\eta_{(n)}$'s are called transverse viscosity since they are transverse to $b_{i}b_{j}$.

In order to find the five shear viscosity coefficients, we first note that the definition of shear stress from 
kinetic theory is 
\begin{equation}
\label{eq:shStrssKT}
    \sigma^{ij}=-\frac{g}{\left(2\pi\right)^3}\int \delta f({\bf p})   v^{i} p^{j} d^3p
\end{equation}
where  $g$ is degeneracy,${\bf v}$ is the particle velocity, and ${\bf p}$ is the corresponding momentum.
The correction to the equilibrium distribution function $\delta f({\bf p})$ can be written as 
\begin{equation}
    \label{eq:deltaf}
    \delta f(p) = -\left(\frac{\partial f_{0}}{\partial\epsilon}\right)C_{ijkl}(\epsilon)v^{i}p^{j}V^{kl},
\end{equation}
where $\epsilon$ is the energy. Hitherto unknown fourth-rank tensor $C_{ijkl}(\epsilon)$ needs to be 
evaluated in order to determine $\delta f$. We note that $C_{ijkl}(\epsilon)$ is contracted with 
symmetric tensors $V^{kl}$ and $v^{i}p^{j}$ to make $\delta f$ scalar, and this imply the following 
symmetry properties 
\begin{equation}
\label{eq:CtensorProp}
    C_{ijkl}=C_{jikl}=C_{ijlk}.
\end{equation}

The fourth rank tensor $C_{ijkl}$ can be constructed from the linear combinations of
$b_{i}$, $\delta_{ij}$, and $\epsilon_{ijk}$. We can construct following eight linearly 
independent fourth rank tensor \cite{Landau:book10}
\begin{eqnarray}
\label{eq:XI1}
 \xi^{(1)}_{ijkl}&=&\delta_{ik}\delta_{jl}+\delta_{il}\delta_{jk},\\
 \xi^{(2)}_{ijkl}&=&\delta_{ij}\delta_{kl}, \\
 \xi^{(3)}_{ijkl}&=&\delta_{ik}b_{j}b_{l}+\delta_{jk}b_{i}b_{l}+\delta_{il}b_{j}b_{k}+\delta_{jl}b_{i}b_{k}, \\
 \xi^{(4)}_{ijkl}&=&\delta_{ij}b_{k}b_{l},\\
 \xi^{(5)}_{ijkl}&=&b_{i}b_{j}\delta_{kl}, \\
 \xi^{(6)}_{ijkl}&=&b_{i}b_{j}b_{k}b_{l}, \\
 \xi^{(7)}_{ijkl}&=&b_{ik}\delta_{jl}+b_{jk}\delta_{il}+b_{il}\delta_{jk}+b_{jl}\delta_{ik}, \\
 \label{eq:XI8}
 \xi^{(8)}_{ijkl}&=&b_{ik}b_{j}b_{l}+b_{jk}b_{i}b_{l}+b_{il}b_{j}b_{k}+b_{jl}b_{i}b_{k}.
\end{eqnarray}
The $C_{ijkl}$ is linear combination of the above fourth rank tensors 
\begin{equation}
\label{eq:CnExp}
    C_{ijkl}(\epsilon)=\tau_{c}\sum_{n=1}^{8} c^{(n)}(\epsilon)\xi^{(n)}_{ijkl}.
\end{equation}

Now let us substitute $\delta f$ from Eq.(\ref{eq:deltaf}) into Eq.(\ref{eq:shStrssKT}) which yields 

\begin{equation}
\label{eq:shStrssKT2}
    \sigma^{ij}=-\frac{g}{\left(2\pi\right)^3}\int \left(\frac{\partial f_{0}}{\partial\epsilon}\right)C_{klmn}(\epsilon)v^{k}p^{l}v^{i} p^{j} d^3pV^{mn}.
\end{equation}

Comparing Eq.(\ref{eq:shStrssKT2}) with Eq.(\ref{eq:stressTens}) we have
\begin{equation}
\label{eq:Eta4thrank1}
    \eta^{ijkl}= -\frac{g}{\left(2\pi\right)^3}\int \left(\frac{\partial f_{0}}{\partial\epsilon}\right)C^{klmn}(\epsilon)v_{m}p_{n}v^{i} p^{j} d^3p.
\end{equation}

Since $v^{i}=\frac{p^{i}}{\epsilon}$, where $\epsilon$ is the energy of the particle, 
the above equation can be re-written as 
\begin{equation}
\label{eq:Eta4thrank2}
    \eta^{ijkl}= -\frac{g}{\left(2\pi\right)^3}\int \left(\frac{\partial f_{0}}{\partial\epsilon}\right)C^{klmn}(\epsilon)\frac{p^{i} p^{j}p_{m}p_{n}}{ \epsilon^{2}}d^3p.
\end{equation}

The product $p^{i}p^{j}p^{m}p^{n}$ on the right hand side of Eq.(\ref{eq:Eta4thrank2}) can be written as  
\begin{equation}
\label{eq:fourPTensor}
   p^{i}p^{j}p^{m}p^{n}=\frac{1}{15}\left(\delta^{ij}\delta^{mn}+\delta^{im}\delta^{jn}+\delta^{in}\delta^{jm}\right)p^{4}.
\end{equation}
The  normalization constant $\frac{1}{15}$ is obtained by contracting the index $i$ with $j$ and $m$ with $n$ (note the superscript
and subscript have the same meaning since we are dealing with three dimensional vectors).
Using Eq.(\ref{eq:fourPTensor}) in Eq.(\ref{eq:Eta4thrank2}) we have 
\begin{equation}
\label{eq:etaTens}
    \eta^{ijkl}= -\frac{1}{15}\frac{g}{\left(2\pi\right)^3}\int \left(\frac{\partial f_{0}}{\partial\epsilon}\right)\frac{p^{4}}{\epsilon^{2}}D^{ijkl}d^3p,
\end{equation}

where 
\begin{equation}
\label{eq:CnDn}
    D^{ijkl}=\left(\xi_{(1)}^{ijmn}+\xi_{(2)}^{ijmn}\right)C^{mnkl}(\epsilon).
\end{equation}

In order to find $C^{ijkl}$ and hence $\delta f$ let us substitute Eq.(\ref{eq:deltaf}) and Eq.(\ref{eq:XI1}-\ref{eq:XI8}) in the Boltzmann equation
Eq.(\ref{eq:Boltzman2}) 
\begin{equation}
\label{eq:CnCoeff}
    \left(\xi^{(1)}_{ijkl}+\chi_{H}\xi^{(7)}_{ijkl} \right)C_{klmn}=\tau_{c}\left(\xi^{(1)}_{ijmn}-\frac{2}{3}\xi^{(2)}_{ijmn}\right).
\end{equation}
Here $\chi_{H}=\frac{qB}{\epsilon}\tau_{c}$ is a dimensionless quantity also known as Hall parameter. 
Using the expansion Eq.(\ref{eq:CnExp}) in Eq.(\ref{eq:CnCoeff}) and taking appropriate tensor contraction 
one can evaluate $c^{(n)}(\epsilon)$'s with the help of Table~\ref{table:1}. It is a straight forward but tedious 
calculation part of which we discuss in the next section and in appendix \ref{App:CnDn}.
The shear viscosity coefficients can be calculated once we know the $D^{ijkl}$ given 
in Eq.(\ref{eq:CnDn}). This can be done in a similar way like the evaluation of $c_{(n)}$'s
and also discussed later in the text and in Appendix \ref{App:CnDn}.

\subsection{$\delta f$ correction due to magnetic field}

As mentioned earlier, in order to evaluate the $\delta f$ we need to find the unknown coefficient 
$c^{(n)}$'s in Eq.(\ref{eq:CnExp}). This can be achieved by using Eq.(\ref{eq:CnExp}) in Eq.(\ref{eq:CnCoeff}) 
and taking the appropriate inner product on both sides. Alternatively one can also
evaluate $c^{(n)}$'s from Eq.(\ref{eq:CnDn}) by first evaluating $D^{ijkl}$ (discussed in the next section). 
Here we solve Eq.(\ref{eq:CnCoeff}) and obtain the following values of $c^{(n)}$'s  (for details see Appendix \ref{App:CnDn})

\begin{figure}
    \resizebox{0.4\textwidth}{!}{%
  \includegraphics{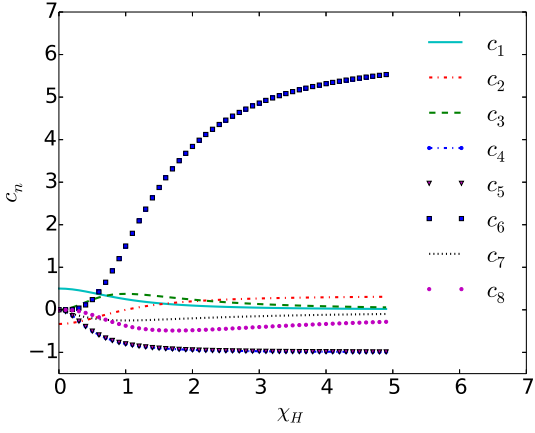}}
    \caption{The coefficients $c_{n}$ as a function of $\chi_{H}$.}
      \label{fig:CnsVsChi}
\end{figure}
\begin{eqnarray}
 \nonumber
 c^{(1)}&=&\frac{1}{2(1+\chi_H^2)}, \\
 \nonumber
 c^{(2)}&=&-\frac{(1-\chi_H^2)}{3(1+\chi_H^2)},\\
 \nonumber
 c^{(3)}&=&\frac{3\chi_H^2}{2(1+\chi_H^2)^2}, \\
 \nonumber
 c^{(4)}&=&c^{(5)}=\frac{-4\chi_H^2}{1+4\chi_H^2}, \\
 \nonumber
 c^{(6)}&=&\frac{6\chi_H^4}{(1+\chi_H^2)^2}, \\
 \nonumber
 c^{(7)}&=&\frac{-\chi_H}{2(1+\chi_H^2)}, \\
 \label{eq:cn}
 c^{(8)}&=&\frac{-3\chi_H^3}{2(1+\chi_H^2)^2}.
\end{eqnarray}
The correction to distribution function $\delta f$ is now readily obtained as
\begin{equation}
    \label{eq:deltafFinal}
    \delta f(p) = -\tau_{c}\sum_{n=1}^{8} c_{(n)}\xi^{(n)}_{ijkl}\left(\frac{\partial f_{0}}{\partial\epsilon}\right)v^{i}p^{j}V^{kl}.
\end{equation}
Here we note that $c^{(4)}=c^{(5)}$, also the coefficients $c^{(7)}$ and $c^{(8)}$ depends linearly 
and third power on the Hall coefficient $\chi_{H}=\frac{qB}{\epsilon}\tau_{c}$, all other $c^{(n)}$'s contain 
quadratic or even power of $\chi_{H}$. The implication is that $c^{(7)}$ and $c^{(8)}$ are electric charge
dependent and hence the $\delta f$. The dependence of $c^{(n)}$ on $\chi_{H}$ is shown in figure \ref{fig:CnsVsChi}.
It will be interesting to investigate in future the effect of the
magnetic field at the freezeout on the charge dependent elliptic flow when the  $\delta f$ correction (Eq.(\ref{eq:deltafFinal})) is used in 
the Cooper-Frye prescription for calculating the invariant yield. We leave this detail investigation for a possible future work 
and consider (in a later section) a simple Bjorken expansion of fluid to study the effect of $\delta f$ 
on transverse momentum spectra of pion.


\subsection{shear viscosity in magnetic field}

Here we discuss calculation of $\eta_{(n)}$'s, for that we shall first determine the equation
for $D^{ijkl}$. Using the definition of $D^{ijkl}$ from Eq. (\ref{eq:CnDn}) in Eq.(\ref{eq:CnCoeff})
and using table~\ref{table:1} of Appendix~\ref{App:CnDn} we have 
\begin{equation}
\label{eq:EqForD}
     \left(\xi^{(1)}_{ijkl}-\frac{2}{5}\xi^{(2)}_{ijkl}+\chi_{H}\xi^{(7)}_{ijkl} \right)D_{klmn}=2\tau_{c}\left(\xi^{(1)}_{ijmn}-\frac{2}{3}\xi^{(2)}_{ijmn}\right).
\end{equation}
 Let us express $D^{ijkl}$ in terms of unknown coefficients 
$d^{(1)}(\epsilon),\allowbreak ...,d^{(8)}(\epsilon)$ and $\tau_{c}$ as
\begin{equation}
\label{eq:Ddecomp}
    D_{ijkl}(\epsilon)=\tau_{c} \sum_{n=1}^{8} d^{(n)}(\epsilon)\xi^{(n)}_{ijkl}. 
\end{equation}
Solving Eq.(\ref{eq:EqForD}) we obtain the values of $d^{(n)}(\epsilon)$ (the details of which is given in
the Appendix \ref{App:CnDn})
\begin{eqnarray}
\nonumber
    d^{(1)}&=&\frac{1}{1+4\chi_H^2}, \\
    \nonumber
    d^{(2)}&=&-\frac{2}{3}\frac{(1-2\chi_H^2)}{(1+4\chi_H^2)},\\
    \nonumber
    d^{(3)}&=&\frac{3\chi_H^2}{(1+4\chi_H^2)(1+\chi_H^2)}, \\
    \nonumber
    d^{(4)}&=&d^{(5)}=\frac{-4\chi_H^2}{1+4\chi_H^2},\\
   \nonumber
    d^{(6)}&=&\frac{12\chi_H^4}{(1+4\chi_H^2)(1+\chi_H^2)},\\
    \nonumber
    d^{(7)}&=&\frac{-\chi_H}{1+4\chi_H^2}, \\
    \label{eq:DnCoeff}
    d^{(8)}&=&\frac{-3\chi_H^3}{(1+4\chi_H^2)(1+\chi_H^2)}.
\end{eqnarray}
Similar to $c^{(n)}$ here we also observe $d^{(4)}=d^{(5)}$.
Now we can calculate the $\eta_{1}..\eta_{5}$ from Eq.(\ref{eq:Eta4thrank2}) by using these $d^{(n)}$'s.
In order to do that we use the definition of traceless shear tensor $\sigma^{ij}$ given in Eq.(\ref{eq:etanSn}), 
and also use Eq.(\ref{eq:trlessShr1}-\ref{eq:trlessShr5}) and Eq.(\ref{eq:XI1}-\ref{eq:XI8}) to express $\eta^{ijkl}$ in 
terms of $\xi^{ijkl}_{(n)}$ as 
\begin{eqnarray}
\nonumber
\label{eq:EtaDecomp}
    \eta^{ijkl}&=&\eta_{(0)}\left(3\xi_{(6)}-\xi_{(5)}-\xi_{(4)}+\xi_{(2)}/3\right)^{ijkl} \\
    \nonumber
               &+&\eta_{(1)}\left(\xi_{(1)}-\xi_{(2)}-\xi_{(3)}+\xi_{(4)}+\xi_{(5)}+\xi_{(6)}\right)^{ijkl}\\
               \nonumber
                &+&\eta_{(2)}\left(\xi_{(3)}-4\xi_{(6)}\right)^{ijkl}\\
                \nonumber
                &-&\eta_{(3)}\left(\xi_{(7)}/2-\xi_{(8)}/2\right)^{ijkl}\\
                 &-&\eta_{(4)}\left(\xi_{(8)}\right)^{ijkl}.
\end{eqnarray}
    
\begin{figure}
    \centering
    \resizebox{0.4\textwidth}{!}{%
  \includegraphics{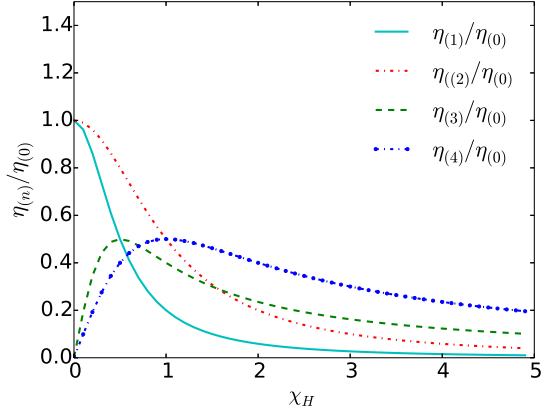}}
    \caption{Ratio of $\eta_{(n)}/\eta_{(0)}$ as a function of $\chi_{H}$.}
    \label{fig:shear}
\end{figure}

Using the above expansion of $\eta^{ijkl}$ on the left hand side of Eq.(\ref{eq:etaTens}) and contracting both side 
with the appropriate tensors we obtain  $\eta_{(n)}$. The straightforward but tedious calculation of 
$\eta_{(n)}$'s are given in Appendix~\ref{AppB}, here we only note the values of $\eta_{(n)}$
\begin{eqnarray}
\label{eq:eta1}
\nonumber
\eta_{(1)}&=&\frac{1}{1+4\chi_{H}^{2}}\eta_{(0)},\\
\nonumber
\eta_{(2)}&=&\frac{1}{1+\chi_{H}^{2}}\eta_{(0)},\\
\nonumber
\eta_{(3)}&=&\frac{2\chi_{H}}{1+4\chi_{H}^{2}}\eta_{(0)},\\
\label{eq:eta4}
\eta_{(4)}&=&\frac{\chi_{H}}{1+\chi_{H}^{2}}\eta_{(0)}.
\end{eqnarray}
Here we observe that $\eta_{(3)}=2\chi_{H}\eta_{(1)}$ and $\eta_{(4)}=\chi_{H}\eta_{(2)}$. 
The variation of $\eta_{(j)}/\eta_{(0)}$ (where $j=1-4$) as a function of $\chi_{H}$ is shown in figure~\ref{fig:shear}.
One still need to evaluate the longitudinal shear viscosity $\eta_{(0)}$.
Once $\eta_{(0)}$ is known for a given system, the rest of the $\eta_{(n)}$'s are obtained from 
Eq.(\ref{eq:eta1}) for a given value of $\chi_{H}$. We shall evaluate $\eta_{(0)}$ for a 
few cases.

\subsection{calculation of $\eta_{(0)}$ for Bose and Fermi gas}

\begin{figure}
    \centering
    \resizebox{0.35\textwidth}{!}{%
  \includegraphics{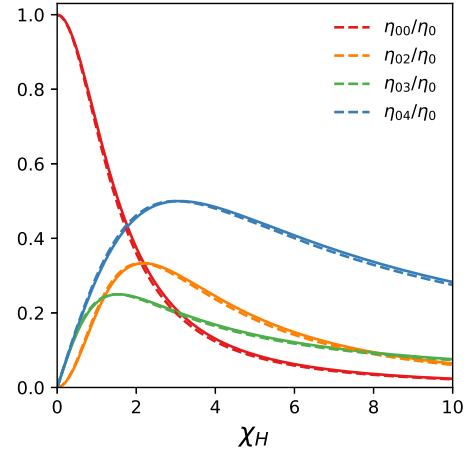}}
    \caption{Comparison of shear viscosity coefficients obtained from Grad's moment method (dotted lines) taken from \cite{Denicol:2018rbw} and our result (solid lines) where we scaled $\chi_H$ for our results by a constant factor.}
    \label{fig:EtaCompare}
\end{figure}
The expression for longitudinal shear viscosity as evaluated in Appendix \ref{AppB} (Eq.(\ref{eq:Apndxeta0})) is 
\begin{equation}
\label{eq:eta0}
    \eta^{(0)} = - \frac{1}{15}\frac{g\tau_{c}}{(2\pi)^3}\int \left(\frac{\partial f_0}{\partial \epsilon}\right)
              \frac{p^4}{\epsilon^2}d^3p.
\end{equation}
For massless Bose and Fermi gas $f_{0}=\frac{1}{exp\left(\left(\epsilon-\mu\right)/T\right)\mp 1}$, 
where $\epsilon=p$ and $\mu$ is the chemical potential.
We note that for this case $\frac{\partial f_{0}}{\partial \epsilon}=-\frac{\partial f_{0}}{\partial \mu}$ 
and Eq.(\ref{eq:eta0}) can be expressed as 
\begin{equation}
\label{eq:eta01}
    \eta_{(0)}=\frac{1}{15}\frac{g\tau_{c}}{(2\pi)^3}\frac{\partial}{\partial \mu}\int \frac{p^4}{{\epsilon}^2}f^0 d^{3}p.
\end{equation}
Eq.(\ref{eq:eta01}) is a special case of the following general form 
\begin{equation}
\label{eq:generalInte}
    B_{n}=\int_0^{\infty}\frac{x^{n} d x}{\exp(x-a) \mp 1}=\pm \Gamma(n+1)Li_{1+n}(\pm e^{a}),
\end{equation}
where $a$ is a constant, $Li_n(x)$ is the poly-logarithmic function of order $n$, and $\Gamma(n)$ is the gamma function.
Using the general result Eq.(\ref{eq:generalInte}) in Eq.(\ref{eq:eta01}) yields 
\begin{equation}
 \eta_{(0)}=\frac{4g\tau_{c} Li_4(\pm e^{\mu/T})}{5{\pi}^2}T^4,
\end{equation}
where $\pm$ corresponds to Bose and Fermi gas respectively. The $\eta_{0}$ at $\mu=0$ for bosons
and fermions are $\eta_{(0)}=\frac{2}{225}g\tau_{c}{\pi}^2T^4$, and $\eta_{(0)}=\frac{7}{900}g\tau_{c}{\pi}^2 T^4$
respectively.

Now let us discuss the case for non-zero mass. 
It is not possible to evaluate the integral Eq.(\ref{eq:eta0}) in a closed form for an arbitrary mass of 
the particles, however, for $m\ll T$, the integral in Eq.(\ref{eq:eta0}) can be expanded in terms of $m/T$, 
and we have
\begin{equation}
\eta_{(0)}=\frac{8}{150}\frac{g\tau_{c}T^4}{{\pi}^2} \left(24 Li_4(\pm e^{\mu/T})-5 Li_2(\pm e^{\mu/T}){\left(\frac{m}{T}\right)}^2+...\right)
\end{equation}
where again the symbol $\pm$ corresponds to bosons or fermions respectively.


\subsection{Comparison of shear viscous coefficients in relaxation time and Grad's 14 moment approximation}
Recently, \cite{Denicol:2018rbw} calculated the shear viscous coefficients in the Grad's 14 moment approach and here we give a brief comparison between the coefficients
calculated in that approach and the relaxation time approach used in this work. Since, \cite{Denicol:2018rbw} used a different basis for the decomposition of the viscous stress tensor $S^{ij}$, we first enumerate the relationships between the $\eta_{(i)}$ defined in this work and $\eta_{ii}$ of \cite{Denicol:2018rbw}. The relationships can be found by noting that
$\eta_1$, $\eta_3$ are the viscosity coefficients for which the corresponding
 shear-stress tensor satisfy the following relationship: $b_iS^{ij}=b_jS^{ij}=0$ and 
 for $\eta_2$, $\eta_4$ the corresponding relationship is: $b_{i}b_{j}S^{ij}=0$, 
 where $S^{ij}$'s are defined in Eqs.(\ref{eq:trlessShr1}-\ref{eq:trlessShr5}).
 Similar calculation can also be found in \cite{Huang:2009ue}.  

The relationship between the $\eta_{0i}$ of \cite{Denicol:2018rbw} and $\eta_{i}$ 
of the present work are given as:
\begin{eqnarray}
\label{eq:relationeta}
\eta_{00}(\xi)&=&\eta_{(1)}(\chi_H)\nonumber\\
\eta_{02}(\xi)&=&\eta_{(2)}(\chi_H)-\eta_{(1)}(\chi_H)\nonumber\\
\eta_{03}(\xi)&=&\eta_{(3)}{(\chi_H)}/{2}\nonumber\\
\eta_{04}(\xi)&=&\eta_{(4)}(\chi_H)
\end{eqnarray}
where $\xi=\frac{qB}{T}\lambda(\propto\tau_c)$, is a dimensionless variable similar to the Hall coefficient $\chi_H$ and $T$ is the temperature. We have not included in the comparision, the term $\eta_{01}(\xi)=\frac{3}{4}(\eta_{(1)}(\chi_H)+\frac{1}{2}\zeta_{(1)}(\chi_H)-\frac{3}{2}\zeta(\chi_H))$, where $\zeta$ is the usual bulk viscosity and $\zeta_{1}$ is the modified bulk viscous coefficient in presence of a magnetic field, since the calculation of bulk viscosity has not been carried out in this work. In figure \ref{fig:EtaCompare}  we show the comparison between $\eta_{0i}$ and the linear combination of $\eta_{(i)}$ as given in Eq.(\ref{eq:relationeta}). As one can see from figure \ref{fig:EtaCompare} Grad's moment method and relaxation time approximation give identical
results when we scale $\chi_H$ appropriately as is done here.


\section{Effect of $\delta f$ to the transverse momentum spectra }
\subsection{In a boost invariant one dimensional Bjorken expansion}


\begin{figure}
    \centering
     \resizebox{0.4\textwidth}{!}{%
   \includegraphics{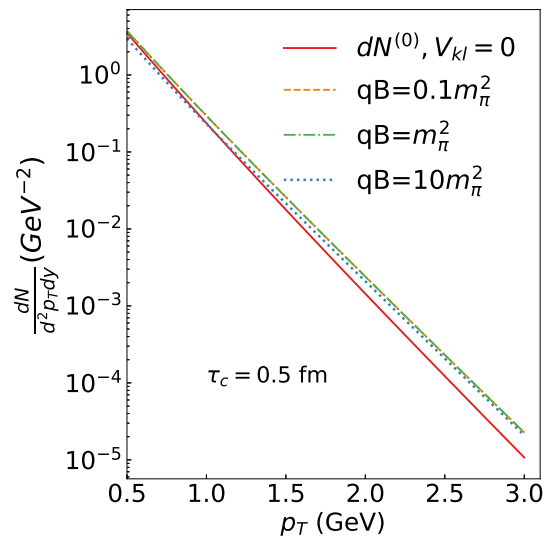}}
    \caption{Invariant yields of $\pi^{-}$ as a function of $p_{T}$ . Solid red line corresponds to without magnetic field and zero shear stress ($V_{kl}=0$) and other lines correspond to different values of magnetic field. }
    \label{fig:Bjspec_trans}
\end{figure}


\begin{figure}
    \centering
    \resizebox{0.4\textwidth}{!}{%
  \includegraphics{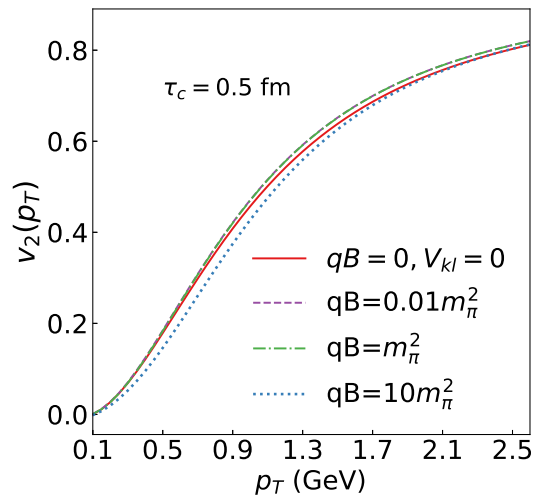}}
    \caption{Elliptic flow $v_{2}$ of $\pi^{-}$ as a function of $p_{T}$ . Solid red line corresponds to without magnetic field and zero shear stress ($V_{kl}=0$) and other lines correspond to different values of magnetic field. }
    \label{fig:BjTransv2}
\end{figure}

One needs to implement the $\delta f$ correction in the Cooper-Frye freezeout formula in 
a numerical magnetohydrodynamics code to investigate the actual effect of $\delta f$ on
invariant yield and flow coefficients. Such a involved study is out of the scope of the 
present work. For the sake of simplicity, here we consider a Bjorken expansion of fluid and 
calculate the corresponding invariant yield of pion in a magnetic field by 
using the distribution function obtained in Eq.(\ref{eq:deltafFinal}). 
The fluid four velocity in this case takes the following
form $(u^{\tau}=1,u^{\eta_{s}}=u^{r}=u^{\phi}=0)$, where we define the longitudinal 
proper time $\tau=\sqrt{t^2-z^2}$, space-time rapidity 
$\eta_{s}=\frac{1}{2}\ln\left(\frac{t+z}{t-z}\right)$, $r=\sqrt{x^2+y^2}$ and 
$\phi=\arctan(y/x)$. The only non vanishing component of the shear stress 
for this velocity profile is $V_{kk}=V_{\eta_{s}\eta_{s}}=-\frac{1}{\tau}$ \cite{Teaney:2003kp}.
When the above value of $V_{\eta_{s}\eta_{s}}$ is used in Eq.(\ref{eq:deltafFinal}) the terms which 
are non-vanishing, contains magnetic field only along the $\eta_{s}$ direction. 
Hence we are forced to choose a hypothetical external magnetic field which 
is strongest along the $\eta_{s}$ direction (0,0,0,$B_{\eta_{s}}$). 
In reality we know that the magnetic field is strongest in the transverse plane and 
vanishing along longitudinal direction.
\begin{figure}
     \resizebox{0.4\textwidth}{!}{%
  \includegraphics{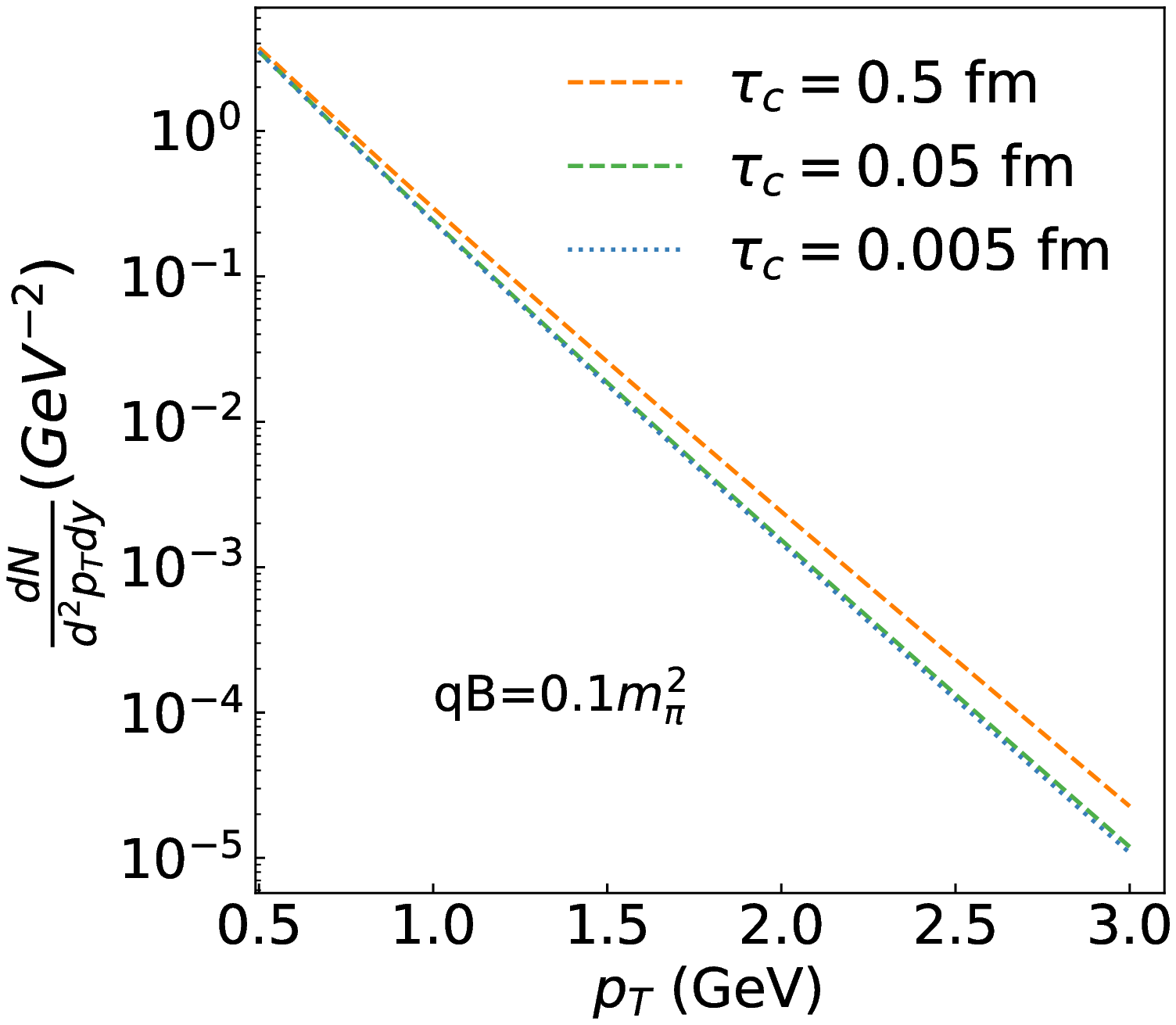}}
  \hspace*{0.1in} 
  \resizebox{0.4\textwidth}{!}{%
  \includegraphics{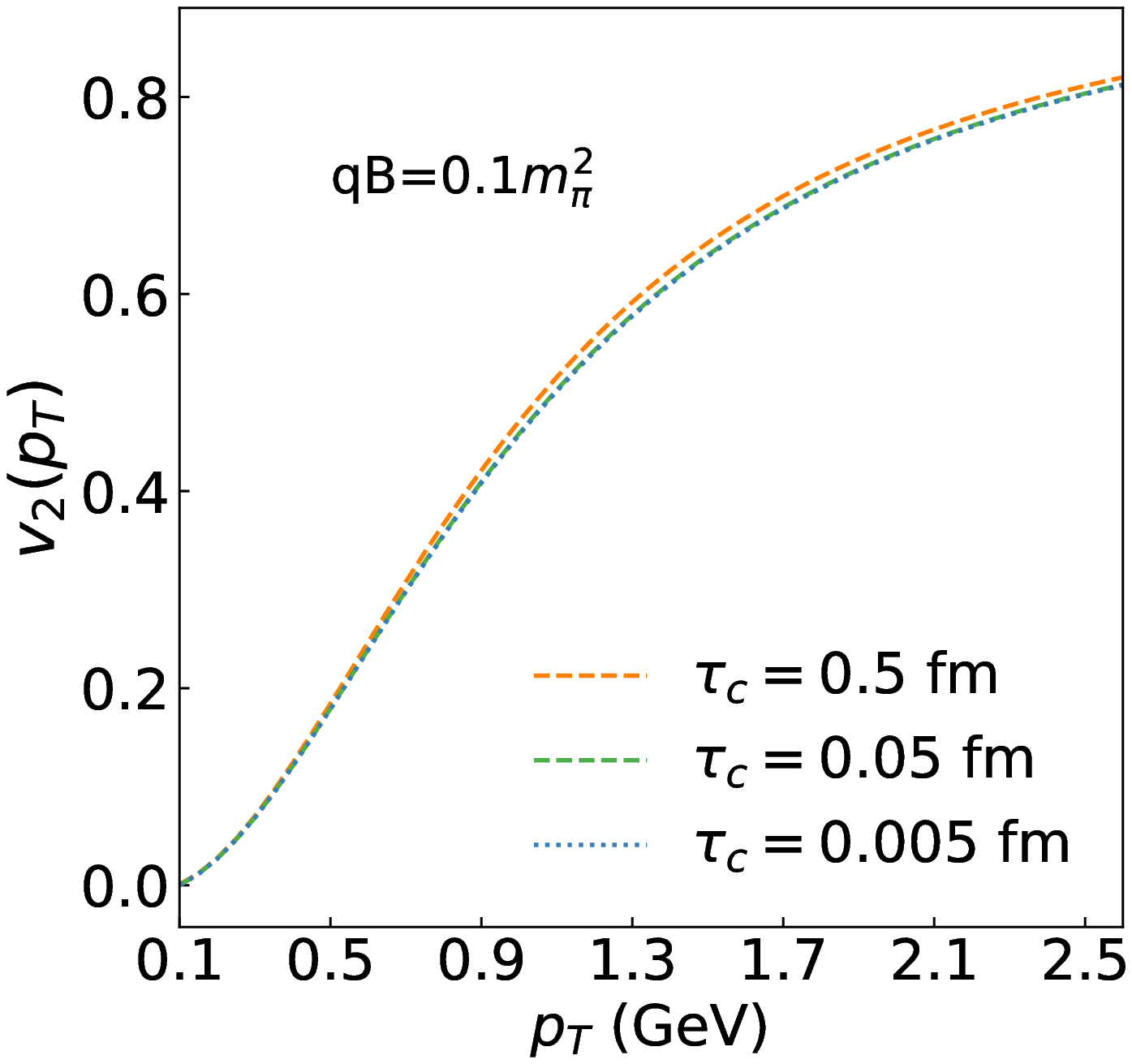}}
    \caption{Top panel: Invariant yield of $\pi^{-}$ as a function of $p_{T}$ or various values of relaxation time when magnetic field kept constant $qB=0.1m_{\pi}^{2}$. Bottom panel: same as top panel but for elliptic flow $v_{2}$ of $\pi^{-}$. }
    \label{fig:DiffRel}
\end{figure}
The invariant yield of pions are obtained from the Cooper-Frye freezeout formula which for the 
present case for a Boltzmann gas is 
\begin{eqnarray}
\label{eq:Spec0}
\frac{d^2N^{(0)}}{d^2p_Tdy}+\sum_{i=1}^{8}\frac{d^2N^{(i)}}{d^2p_Tdy}&=&\frac{1}{(2\pi)^{3}}\int{p^{\mu}d\Sigma_{\mu}}\left(f_{0}+\sum_{i=1}^{8}\delta f^{(i)}\right),\nn\\
\end{eqnarray} 
where $x=m_{T}/T$, modified Bessel function of second kind  $K_{\nu}(x)$
the differential freezeout hypersurface $d\Sigma^{\mu}=\left(\tau d\eta_{s} rdr d\phi,0,0,0 \right)$, and the 
four momentum $p^{\mu}=$\newline $ (m_{T}\cosh y ,p_{T}\cos\phi,p_{T}\sin\phi,m_{T}\sinh y)$. The superscript $0$ in Eq.(\ref{eq:Spec0})
indicates the equilibrium case. The first integral on the right hand side is  
$\frac{2m_T\pi R_0^2\tau_{0}}{(2\pi)^{3}} K_1(x)$ and the second integral is the correction to the invariant yield.
We evaluate the correction by using the appropriate form of $\delta f^{(i)}$'s (given in 
appendix \ref{App:3}) in Eq.(\ref{eq:Spec0}) and evaluating the integral numerically. 
For the calculation we have also used $\chi_{H}(\epsilon)=(q B \tau_{c})/(m_{T}\cosh y)$, $\tau_{c}=$ 0.5 fm, 
and we take midrapidity i.e., $y=0$.  We found that the $p_{T}$ spectra hardly changes in magnetic field compared to
 the without magnetic field case even for a ambitious value of $qB\sim 100 m_{\pi}^{2}$. 
 For a more realistic fluid expansion which contains 
 transverse as well as azimuthal variation of fluid velocity, the correction 
 to the $p_T$ spectra and flow coefficients might be quite different than what is obtained here, this case is
 discussed in the next section.
 It is also worthwhile to mention that $\delta f$ contains two terms $\delta f^{(7)}$ and $\delta f^{(8)}$ which are charge dependent
 ($c^{(7)},c^{(8)}$ in Eq.(\ref{eq:cn})) which might be important for studying CME.  
 For the present case of one dimensional Bjorken expansion these charge dependent terms vanishes.  


\begin{figure}
    \centering
    \resizebox{0.4\textwidth}{!}{%
  \includegraphics{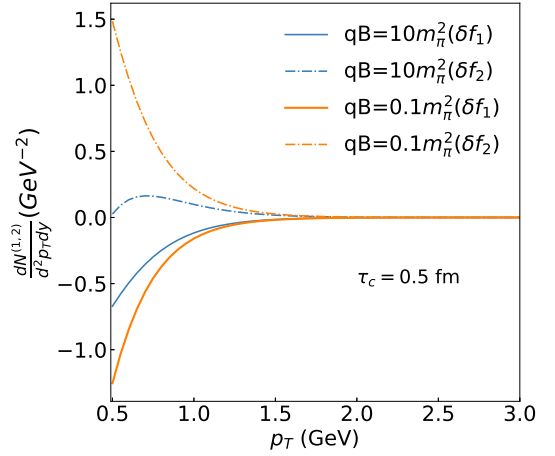}}
    \caption{Invariant yield of $\pi^{-}$ corresponding to $\delta f_{1}$ and $\delta f_{2}$ as a function of $p_{T}$ 
     for two different magnetic fields.  Orange line corresponds to $qB=0.1 m_{\pi}^{2}$ and blue line corresponds 
     to $qB=10m_{\pi}^{2}$. }
    \label{fig:DeltaFbyF}
\end{figure}

\subsection{Transverse expansion along with the longitudinal Bjorken expansion}
\label{subsec:TransExpns}
In order to study a more realistic fluid expansion with non-zero transverse flow, following \cite{Teaney:2003kp} 
we use a generalise version of the blast wave model.  We would like to mention that the blast 
wave model is a simple model of the flow fields and 
 a full dissipative magnetohydrodynamics simulation is needed to estimate actual effects.
By assuming a linear rise of transverse velocity as a function of radius of the fireball in the 
transverse plane and a velocity field with a small elliptic flow component we have the 
following hydrodynamical fields :

\begin{eqnarray}
\label{eq: transFlow}
T(\tau_0,\eta_s,r,\phi)&=&T_0\Theta(R_0-r),\nonumber\\
u^{r}(\tau_0,\eta_s,r,\phi)&=&u_0\frac{r}{r_0}(1+u_2\cos(2\phi))\Theta(R_0-r),\nonumber\\
u^{\phi}&=&0,\nonumber\\
u^{\eta}&=&0,\nonumber\\
u^{\tau}&=&\sqrt{1+(u^r)^2}.\nonumber\\
\end{eqnarray}

For a head-on collision the elliptic flow component $u_2=0$. 
Here we have taken, $u_{0}=0.55$ and $u_{2}=0.1$. These values 
approximately corresponds to a mid central heavy ion collisions at top
RHIC energies.
It is useful to realise that $\tau u^{\eta}$ and $r u^{\phi}$ are velocities 
in $\eta$ and $\phi$ directions respectively. For boost invariant flow 
$u^{\eta}=0$ and for  rotationally invariant flow $u^{\phi}=0$.
Here we also assume longitudinal boost-invariance, the invariant yield 
is calculated by using Eq.(\ref{eq:Spec0}) for the given hydrodynamic fields 
Eq.(\ref{eq: transFlow}) . Calculation of corresponding $V_{kl}$ are given 
in Appendix \ref{App:TransFlow}.  The invariant yield of $\pi^{-}$
for various values of magnetic fields are shown in figure \ref{fig:Bjspec_trans}.
For comparison we also show the zero magnetic field  and zero shear stress 
case in the same figure by the solid red line. Unlike one dimensional expansion we found visible correction to 
the invariant yield when non-zero transverse expansion is taken into account (note that 
$qB=0$ does not imply ideal case, one also needs to set zero shear viscosity which 
is achieved by setting $V_{kl}=0$).
In order to calculate the $v_{2}$ we use the following formula which is obtained by 
considering the correction to be small (see \cite{Teaney:2003kp})
\begin{eqnarray}
v_{2}\left(p_{T}\right)&=&v_{2}^{0} \left( p_{T} \right) \left(1- \frac{\int d\phi \frac{d^2N^{(i)}}{p_{T}dp_{T}d\phi}}{\int d\phi \frac{d^2N^{(0)}}{p_{T}dp_{T}d\phi}}\right)\nonumber\\
&+&\frac{\int d\phi cos(2\phi) \frac{d^2N^{(i)}}{p_{T}dp_{T}d\phi}}{\int d\phi \frac{d^2N^{(0)}}{p_{T}dp_{T}d\phi}}.
\end{eqnarray}
The $v_{2}$ for different values of magnetic field are shown in figure~\ref{fig:BjTransv2}.
One can clearly see that $v_{2}$ changes due to the $\delta f$ correction in presence of
magnetic field compared to the without magnetic field case (shown by red line in figure~\ref{fig:BjTransv2}).
The above mentioned results are obtained for a constant relaxation time $\tau_{c}=0.5$ fm.
The dependence of invariant yield and $v_{2}$ on $\tau_{c}$ for a fixed value of magnetic 
field $qB=0.1m_{\pi}^{2}$ are shown in figure \ref{fig:DiffRel}.
Finally we would like to comment on the non-monotonic behaviour of $\delta f$
correction as observed in invariant yield and $v_{2}$ (see figure \ref{fig:Bjspec_trans} and 
\ref{fig:BjTransv2} ). We note that the contribution from $\delta f_{1}$ and  $\delta f_{2}$
shows non-monotonic behaviour as a function of magnetic field. This is shown in 
figure \ref{fig:DeltaFbyF} where we plot corresponding corrections in invariant yields 
$\frac{d^{2}N^{(1)}}{d^{2}p_{T}dy}$ and  $\frac{d^{2}N^{(2)}}{d^{2}p_{T}dy}$ respectively 
as function of  $p_{T}$ for two different values of magnetic field $qB=0.1m_{\pi}^{2}$ (orange lines) and 
$qB=10m_{\pi}^{2}$(blue lines). One can clearly see that the corrections due to two terms 
cancel each other for smaller values of magnetic field, whereas for a larger magnetic 
field they act coherently. This behaviour is reflected in the correction to invariant yield and $v_{2}$.

\section{Conclusion}
We calculated the $\delta f$ correction for a fluid with finite velocity gradient and under the influence of
magnetic field in the relaxation time approximation. The $\delta f$
is found to be composed of eight different terms some of which are electric charge dependent. 
When used in the Cooper-Frey freezeout formula for a simple fluid evolution in one dimensional   
Bjorken expansion, the change in invariant yield 
of negative pion is found to be negligible for realistic values of the magnetic field.
Interestingly, however, when the transverse expansion of the fluid is  taken into account
we found that both spectra and $v_2$ shows noticeable changes. 
We also evaluated the shear viscous coefficients in magnetic field. One longitudinal $\eta_{(0)}$ and four transverse
($\eta_{(1)},\eta_{(2)},\eta_{(3)},\eta_{(4)}$) viscous coefficients have been calculated as a 
function of Hall parameter $\chi_{H} \left(=qB\tau_{c}/\epsilon\right)$. It is observed that 
the transverse viscosities are smaller in comparison to the $\eta_{(0)}$ for $0\leq \chi_{H} \leq 5$. It has also been 
shown that $\eta_{(2)}-\eta_{(4)}$ can be expressed in terms of $\eta_{(0)}$ with 
coefficients which are functions of $\chi_{H}$ only. All the transverse
shear viscous coefficients are found to decrease with respect to the $\eta_{0}$ for $\chi_{H}>1$.
At smaller $\chi_{H}\leq 1$, the $\eta_{3}$ and $\eta_{4}$ shows non-monotonic behavior. 
We would like to mention that it is important to calculate the same $\delta f$ in other methods
such as Chapman-Enskog or Grad's moment method in order to corroborate our findings.
For example, it is interesting to compare results from other methods to see whether one also 
get similar electric charge dependent terms in the $\delta f$ corrections which might be important 
for studying CME. These open problems are out of the scope of the present study and we leave it 
for a possible future investigation.
\begin{table*}[t]
\centering
\begin{tabular}{ c| c c c c c c c c}
\hline
 &$\1x$ & $\2x$ & $\3x$ & $\4x$ & $\5x$ & $\6x$ & $\7x$ & $\8x$ \\ [2ex]
\hline
$\1x$ & $2\1x$ & $2\2x$ & $2\3x$ & $2\4x$ & $2\5x$ & $2\6x$ & $2\7x$ & $2\8x$ \\  
 $\2x$ & $2\2x$ & $3\2x$ & $4\4x$ & $3\4x$ & $\2x$ & $\4x$ & 0 & 0 \\  
  $\7x$ & $2\7x$ & 0 & $2\8x$ & 0 & 0 & 0 & $-8(\1x-\2x-\frac{3}{4}\3x+\4x+\5x)$ & $2(4\6x-\3x)$ \\   
 \hline
  \end{tabular}
 \caption{Product of non-commutative basis elements $\xi^{(r)}_{ijkl}*\xi^{(s)klmn}$}
\label{table:1}     
\end{table*} 
\section*{Acknowledgement}
V.R would like to thank Rishi Sharma for helpful discussion. PM and VR are supported by DST-INSPIRE Faculty
research grant, India. We also acknowledge useful discussion with Amaresh Jaiswal. 

\appendix
\section{Calculation of $c^{(n)}$ and $d^{(n)}$'s}
\label{App:CnDn}

Here we discuss in details the calculation of coefficients $c^{(n)}$'s and $d^{(n)}$'s appeared in Eq.(\ref{eq:cn})
and Eq.(\ref{eq:DnCoeff}). First let us calculate $c_{n}$'s from Eq.~(\ref{eq:CnCoeff})

\be
\left(\1x_{ijkl}+\chi_H \7x_{ijkl}\right) C_{klmn}=
\tau_{c}\left(\1x_{ijmn}-\frac{2}{3}\2x_{ijmn}\right) 
\ee
where $C_{ijkl}=\tau_{c}\sum_{n=1}^{8} c^{(n)}(\epsilon)\xi^{(n)}_{ijkl}$
\begin{equation}
 \left(\1x_{ijkl}+\chi_H \7x_{ijkl}\right)\tau_{c} \sum_{n=1}^8 c^{n}(\epsilon) \xi^{(n)}_{klmn}=
 \tau_{c}\left(\1x_{ijmn}-\frac{2}{3}\2x_{ijmn}\right)   
\end{equation}
For convenience let us write $c^{(n)}(\epsilon)$ as $c^{(n)}$, swap the right and left 
hand side of the above equation and writing the tensor index as a common term we have 

\bea
\left(\1x-\frac{2}{3}\2x\right)_{ijmn}&=&
\left(\1x+\chi_H \7x\right)_{ijkl} c^{(1)} \1x_{klmn}\nn\\
&+&\left(\1x+\chi_H \7x\right)_{ijkl} c^{(2)} \2x_{klmn}\nn\\
&+&\left(\1x+\chi_H \7x\right)_{ijkl} c^{(3)} \3x_{klmn}\nn\\
&+&\left(\1x+\chi_H \7x\right)_{ijkl} c^{(4)} \4x_{klmn}\nn\\
&+&\left(\1x+\chi_H \7x\right)_{ijkl} c^{(5)} \5x_{klmn}\nn\\
&+&\left(\1x+\chi_H \7x\right)_{ijkl} c^{(6)} \6x_{klmn}\nn\\
&+&\left(\1x+\chi_H \7x\right)_{ijkl} c^{(7)} \7x_{klmn}\nn\\
&+&\left(\1x+\chi_H \7x\right)_{ijkl} c^{(8)} \8x_{klmn}\nn\\
\eea
Now we need to evaluate the tensor products like $\xi_{ijkl}^{(1)}\xi_{klmn}^{(n)}$
and $\xi_{ijkl}^{(7)}\xi_{klmn}^{(n)}$ which can be found from Eq.(\ref{eq:XI1}-\ref{eq:XI8}).
We have tabulated all such tensor products in table~\ref{table:1}. Using those tabulated
values we have 
 
\bea 
\label{eq:ApCcoef}
\left(\frac{1}{2}\1x-\frac{1}{3}\2x\right)_{ijmn}&=&
\left(c^{(1)}-4\chi_H c^{(7)}\right)\xi^{(1)}_{ijmn}\nn\\
&+&\left(c^{(2)}+4\chi_H c^{(7)}\right)\xi^{(2)}_{ijmn}\nn\\
&+&\left(c^{(3)}+3\chi_H c^{(7)}-\chi_H c^{(8)}\right)\xi^{(3)}_{ijmn} \nn\\
&+&\left(c^{(4)}-4\chi_H c^{(7)}\right)\xi^{(4)}_{ijmn}\nn\\
&+&\left(c^{(5)}-4\chi_H c^{(7)}\right)\xi^{(5)}_{ijmn}\nn\\
&+&\left(c^{(6)}+4\chi_H c^{(8)}\right)\xi^{(6)}_{ijmn}\nn\\
&+&\left(c^{(7)}+\chi_H c^{(1)}\right)\xi^{(7)}_{ijmn}\nn\\
&+&\left(c^{(8)}+\chi_H c^{(3)}\right)\xi^{(8)}_{ijmn}\nn\\
\eea

Equating the coefficients of $\xi^{(n)}_{ijmn}$ on both side of Eq.(\ref{eq:ApCcoef}) we have, 
\begin{eqnarray}
\nonumber
  c^{(1)}-4\chi_H c^{(7)}&=&\frac{1}{2} \\
  \nonumber
  c^{(2)}+4\chi_H c^{(7)}&=&-\frac{1}{3}\\
  \nonumber
  c^{(3)}+3\chi_H c^{(7)}-\chi_H c^{(8)}&=&0 \\
  \nonumber
  c^{(4)}-4\chi_H c^{(7)}&=&0 \\
  \nonumber
  c^{(5)}-4\chi_H c^{(7)}&=&0 \\
  \nonumber
  c^{(6)}+4\chi_H c^{(8)}&=&0 \\
  \nonumber
  c^{(7)}+\chi_H c^{(1)}&=&0  \\
  c^{(8)}+\chi_H d^{(3)}&=&0  
\end{eqnarray}
 
By solving these eight equations, we obtain the values of $c^{(n)}$'s given in Eq.(\ref{eq:cn}).

The calculation for $d^{(n)}$'s proceed in a similar way. From Eq.(\ref{eq:EqForD}) we have 
\begin{equation}
     \left(\xi^{(1)}_{ijkl}-\frac{2}{5}\xi^{(2)}_{ijkl}+\chi_{H}\xi^{(7)}_{ijkl} \right)D_{klmn}=2\tau_{c}\left(\xi^{(1)}_{ijmn}-\frac{2}{3}\xi^{(2)}_{ijmn}\right),
\end{equation}
where
\begin{equation}
    D_{ijkl}(\epsilon)=\tau_{c} \sum_{n=1}^{8} d^{(n)}(\epsilon)\xi^{(n)}_{ijkl}. 
\end{equation}
Thus we have 
\begin{eqnarray}
\nonumber
\left(\1x_{ijmn}-\frac{2}{3}\2x_{ijmn}\right)&=&\left(\1x_{ijkl}-\frac{2}{5}\2x_{ijkl}+\chi_H \7x_{ijkl}\right) d^{(1)} \xi^{(1)}_{klmn}\\
&+&\left(\1x_{ijkl}-\frac{2}{5}\2x_{ijkl}+\chi_H \7x_{ijkl}\right) d^{(2)}\2x_{klmn}\nn\\
&+&\left(\1x_{ijkl}-\frac{2}{5}\2x_{ijkl}+\chi_H \7x_{ijkl}\right) d^{(3)} \3x_{klmn}\nn\\
&+&\left(\1x_{ijkl}-\frac{2}{5}\2x_{ijkl}+\chi_H \7x_{ijkl}\right) d^{(4)} \4x_{klmn}\nn\\
&+&\left(\1x_{ijkl}-\frac{2}{5}\2x_{ijkl}+\chi_H \7x_{ijkl}\right) d^{(5)} \5x_{klmn}\nn\\
&+&\left(\1x_{ijkl}-\frac{2}{5}\2x_{ijkl}+\chi_H \7x_{ijkl}\right) d^{(6)} \6x_{klmn}\nn\\
&+&\left(\1x_{ijkl}-\frac{2}{5}\2x_{ijkl}+\chi_H \7x_{ijkl}\right) d^{(7)} \7x_{klmn}\nn\\
&+&\left(\1x_{ijkl}-\frac{2}{5}\2x_{ijkl}+\chi_H \7x_{ijkl}\right) d^{(8)} \8x_{klmn}.\nn\\
\end{eqnarray}
 
Using the tabulated values (table \ref{table:1}) for the tensor product and using the shorthand notation for tensor indices we have

\begin{eqnarray}
\left(\xi^{(1)}-\frac{2}{3}\xi^{(2)}\right)_{ijmn}&=&
d^{(1)}\left(2\xi^{(1)}-\frac{4}{5}\xi^{(2)}+2\chi_{H}\xi^{(7)}\right)_{ijmn}\nn\\
&+&d^{(2)}\left(\frac{4}{5}\xi^{(2)}\right)_{ijmn}\nn\\
&+&d^{(3)}\left(2\xi^{(3)}-\frac{8}{5}\xi^{(4)}+2\chi_{H}\xi^{(8)}\right)_{ijmn}\nn\\
&+&d^{(4)}\left(\frac{4}{5}\xi^{(4)}\right)_{ijmn}\nn\\
&+&d^{(5)}\left(2\xi^{(5)}-\frac{2}{5}\xi^{(2)}\right)_{ijmn}\nn\\
&+&d^{(6)}\left(2\xi^{(6)}-\frac{2}{5}\xi^{(4)}\right)_{ijmn}\nn\\
&+&d^{(7)} [2\xi^{(7)}-8\chi_{H} \nn \\
&&\left(\xi^{(1)}-\xi^{(2)}-\frac{3}{4}\xi^{(3)}+\xi^{(4)}+\xi^{(5)}\right)]_{ijmn}\nn\\
&+&d^{(8)}\left(2\xi^{(8)}+2\chi_{H}\left(4\xi^{(6)}-\xi^{(3)}\right)\right)_{ijmn}.\nn\\
\end{eqnarray}
Comparing the coefficients of $\xi^{(n)}_{ijmn}$ on both side of the above equations we have 
\begin{eqnarray}
\nonumber
d^{(1)}-4\chi_{H} d^{(7)} &=& 1 \\
\nonumber
 \frac{2}{5}d^{(1)}-\frac{2}{5}d^{(2)}+\frac{1}{5}d^{(5)}-4\chi_{H} d^{(7)}&=&\frac{2}{3} \\
 \nonumber
 d^{(3)}+3\chi_{H}d^{(7)}-\chi_{H}d^{(8)} &=& 0 \\
 \nonumber
  -\frac{4}{5}d^{(3)}+\frac{2}{5}d^{(4)}-\frac{1}{5}d^{(6)}-4\chi_{H} d^{(7)} &=& 0 \\
  \nonumber
 d^{(5)}-4\chi_{H}d^{(7)} &=& 0 \\
 \nonumber
 d^{(6)}+4\chi_{H}d^{(8)} &=& 0 \\
 \nonumber
  d^{(7)}+\chi_{H} d^{(1)} &=& 0  \\
 d^{(8)}+\chi_{H}d^{(3)} &=& 0.  \\
\end{eqnarray}

 The values of $d^{(n)}$'s (see Eq.\ref{eq:DnCoeff} ) are obtained by solving these eight equations.
 
\section{Calculation of $\eta_{(n)}$}
\label{AppB}
Here we shall discuss the evaluation of longitudinal viscosity $\eta_{(0)}$ for which 
the final expression is given in Eq.(\ref{eq:eta0}).

\begin{table}[h]
 \centering
\begin{tabular}{ c c c c c c c c c}
\hline
 &$\1x$ & $\2x$ & $\3x$ & $\4x$ & $\5x$ & $\6x$ & $\7x$ & $\8x$ \\ [2ex]
\hline
$\1x$ & 24 & 6 & 16 & 2 & 2 & 2 & 0 & 0 \\  
 $\2x$ & 6 & 9 & 4 & 3 & 3 & 1 & 0 & 0 \\  
  $\3x$ & 16 & 4 & 24 & 4 & 4 & 4 & 0 & 0\\   
   $\4x$ & 2 & 3 & 4 & 3 & 1 & 1 & 0 & 0 \\   
    $\5x$ & 2 & 3 & 4 & 3 & 1 & 1 & 0 & 0\\  
     $\6x$ & 2 & 1 & 4 & 1 & 1 &1 & 0 & 0\\   
      $\7x$ & 0 & 0& 0 & 0 & 0 & 0 & 40 & 8\\   
       $\8x$ & 0 & 0& 0 & 0 & 0 & 0 & 8 & 8\\  
        \hline
       \end{tabular}
 \caption{Scalar product $\xi^{(n)}_{ijkl} \xi^{ijkl}_{(n)}$}
\label{table:2}     
\end{table}

As given in Eq.(\ref{eq:EtaDecomp}) the $\eta^{ijkl}$ can be decomposed as 
\begin{equation}
\label{eq:AppEtaDecom}
    \eta^{ijkl}=\sum_{i=0}^4\eta_{(n)} I_{(n)}^{ijkl},
\end{equation} 
where $I_{(n)}^{ijkl}$'s are defined as 
\bea
I_{(0)}^{ijkl}&=&(3\6x-\4x-\5x+\2x/3)^{ijkl},\nn\\
I_{(1)}^{ijkl}&=&(\1x-\2x-\3x+\4x+\5x+\6x)^{ijkl},\nn\\
I_{(2)}^{ijkl}&=&(\3x-4\6x)^{ijkl},\nn\\
I_{(3)}^{ijkl}&=&-\frac{1}{2}(\7x-\8x)^{ijkl},\nn\\
I_{(4)}^{ijkl} &=& -\left(\8x\right)^{ijkl}. \nn\\
\eea

From kinetic theory the $\eta^{ijkl}$ can be expressed in terms of $\delta f$ as
\be
\eta^{ijkl}=-\frac{1}{15}\frac{g}{(2\pi)^3}\int{d^3p \left(\frac{\partial f_0}{\partial \epsilon}\right)}
              \frac{p^4 c^4}{\epsilon^2}D^{ijkl}
\label{etaa}
\ee
In order to find $\eta_{(0)}$ we use the decomposition given in Eq.(\ref{eq:AppEtaDecom}) 
on the left hand side of the above equation and multiply both side by $I^{(0)}_{ijkl}$ yield
\begin{equation}
\label{eq:compare_eta}
\eta_{(0)}I_{(0)}^{ijkl}I^{(0)}_{ijkl} =-\frac{1}{15}\frac{g}{(2\pi)^3}\int{d^3p \left(\frac{\partial f_0}{\partial \epsilon}\right)}
              \frac{p^4 c^4}{\epsilon^2}D^{ijkl} I^{(0)}_{ijkl}.
\end{equation}

In order to evaluate the above integral we need to find  the product $D^{ijkl} I^{(0)}_{ijkl}$.
By using the expression for $D^{ijkl}$ Eq. (\ref{eq:Ddecomp}) we have 
\begin{equation}
    D^{ijkl} I^{(0)}_{ijkl}= \tau_c\sum_{n=1}^8 d_{(n)}\xi_{(n)}^{ijkl}I^{(0)}_{ijkl}.
\end{equation}
For simplicity and to avoid writing repetitively the tensor index $ijkl$ we remove the indices 
in the following equation 

\begin{widetext}
\begin{eqnarray}
\sum_{n=1}^8  d_{(n)}\xi_{(n)}I^{(0)}
&=&d_{(1)}\xi_{(1)} \left(3\6x-\4x-\5x+\frac{\2x}{3}\right)+ d_{(2)}\xi_{(2)}\left(3\6x-\4x-\5x+\frac{\2x}{3}\right) \nn\\
&+& d_{(3)}\xi_{(3)} \left(3\6x-\4x-\5x+\frac{\2x}{3}\right)+ d_{(4)}\xi_{(4)} \left(3\6x-\4x-\5x+\frac{\2x}{3}\right)\nn\\
&+& d_{(5)}\xi_{(5)} \left(3\6x-\4x-\5x+\frac{\2x}{3}\right) + d_{(6)} \xi_{(6)} \left(3\6x-\4x-\5x+\frac{\2x}{3}\right)\nn\\
&+& d_{(7)}\xi_{(7)} \left(3\6x-\4x-\5x+\frac{\2x}{3}\right) +d_{(8)} \xi_{(8)} \left(3\6x-\4x-\5x+\frac{\2x}{3}\right)\nn\\
&=& d_{(1)}\left(3 \xi_{(1)}\6x-\xi_{(1)}\4x+\xi_{(1)}\5x-\xi_{(1)}\frac{\2x}{3}\right)\nn\\
&+&d_{(2)}\left(3 \xi_{(2)}\6x-\xi_{(2)}\4x+\xi_{(2)}\5x-\xi_{(2)}\frac{\2x}{3}\right) \nn\\
&+&d_{(3)}\left(3 \xi_{(3)}\6x-\xi_{(3)}\4x+\xi_{(3}\5x-\xi_{(3)}\frac{\2x}{3}\right)\nn\\
&+&d_{(4)}\left(3 \xi_{(4)}\6x-\xi_{(4)}\4x+\xi_{(4)}\5x-\xi_{(4)}\frac{\2x}{3}\right)\nn\\
&+&d_{(5)}\left(3 \xi_{(5)}\6x-\xi_{(5)}\4x+\xi_{(5)}\5x-\xi_{(5)}\frac{\2x}{3}\right)\nn\\
&+&d_{(6)}\left(3 \xi_{(6)}\6x-\xi_{(6)}\4x+\xi_{(6)}\5x-\xi_{(6)}\frac{\2x}{3}\right)\nn\\
&+&d_{(7)}\left(3 \xi_{(7)}\6x-\xi_{(7)}\4x+\xi_{(7)}\5x-\xi_{(7)}\frac{\2x}{3}\right)\nn\\
&+&d_{(8)}\left(3 \xi_{(8)}\6x-\xi_{(8)}\4x+\xi_{(8)}\5x-\xi_{(8)}\frac{\2x}{3}\right)\nn\\
&=& 4\left[d^{(1)}+\frac{8}{3}d^{(3)}+d^{(5)}+\frac{2}{3}d^{(6)}\right]\nn\\
&=&  4\left[\left(\frac{1}{1+4\chi_H^2}\right)+\frac{8}{3}\left(\frac{3\chi_H^2}{(1+4\chi_H^2)(1+\chi_H^2)}\right)
      + \left(\frac{-4\chi_H^2}{1+4\chi_H^2}\right)+
      \frac{2}{3}\left(\frac{12\chi_H^4}{(1+4\chi_H^2)(1+\chi_H^2)}\right)\right]\nn\\
&=&4\left[\frac{(1+4\chi_H^2)(1+\chi_H^2)}{(1+4\chi_H^2)(1+\chi_H^2)}\nn\right]\\
&=&4.
\end{eqnarray}
In evaluating the tensor contraction in the above calculation we have used values from table(\ref{table:2}).
\end{widetext}

Now let us calculate the product $I_{(0)}^{ijkl}I^{(0)ijkl}$ which appeared on the left hand side of Eq.(\ref{eq:compare_eta}) 
Again we shall use values given in table~(\ref{table:2}) for the following calculation.
\begin{widetext}
\begin{eqnarray}
I_{(0)}^{ijkl}I^{(0)}_{ijkl}&=&\left(3\6x-\4x-\5x+\frac{\xi^{(2)}}{3}\right)^{ijkl}\left(3\6x-\4x-\5x+\frac{\xi^{(2)}}{3}\right)_{ijkl}\nn\\
&=& 9-3-3+1-3+3+1-1-3+1+3-1+1-1-1+1\nn\\
&=& 4.    
\end{eqnarray}
\end{widetext}

Finally using these values in Eq.~(\ref{eq:compare_eta}) we have the expression for $\eta_{(0)}$ as
\begin{equation}
\label{eq:Apndxeta0}  
    \eta_{(0)} = - \frac{1}{15}\frac{g\tau_{c}}{(2\pi)^3}\int{d^3p \left(\frac{\partial f_0}{\partial \epsilon}\right)}
              \frac{p^4 c^4}{\epsilon^2}.      
\end{equation}
Calculation of $\eta_{(1)}..\eta_{(4)}$ are similar and we don't include them here.

\section{$\delta f$ for Bjorken expansion}
\label{App:3}
Here we discuss the $\delta f^{(i)}$ used in Eq.(\ref{eq:Spec0}) for the longitudinal boost invariant case. The shear stress for 
this case is . For the boost invariant expansion without transverse flow, $u_{\tau}=1$ and all other components of 
velocity are zero ($u_{r}=u_{\eta}=u_{\phi}=0$). 
Thus the only non vanishing component of the shear stress $V_{kk}=V_{\eta\eta}=-\frac{1}{\tau}$. 
The corresponding $\delta f^{(i)}$ for the magnetic field (0,0,0,$B_{\eta}$) 
turns out to be of the following form

 \bea
 \delta f^{(1)}&=&-2 c^{(1)}\tau_c\left(\frac{f_0}{T\epsilon\tau}\right)m_T^2 \sinh^2{\eta_s},\nn\\
\delta f^{(2)}&=&-c^{(2)}\tau_c\left(\frac{f_0}{T\epsilon\tau}\right)m_T^2 \sinh^2{\eta_s},\nn\\
\delta f^{(3)}&=&-4c^{(3)}\tau_c\left(\frac{f_0}{T\epsilon\tau}\right)b_{\eta_s}^2m_T^2 \sinh^2{\eta_s},\nn\\
\delta f^{(4)}&=&-c^{(4)}\tau_c\left(\frac{f_0}{T\epsilon\tau}\right)b_{\eta_s}^2m_T^2 \sinh^2{\eta_s},\nn\\
\delta f^{(5)}&=&-c^{(5)}\tau_c\left(\frac{f_0}{T\epsilon\tau}\right)b_{\eta_s}^2m_T^2 \sinh^2{\eta_s},\nn\\
\delta f^{(6)}&=&-c^{(6)}\tau_c\left(\frac{f_0}{T\epsilon\tau}\right)b_{\eta_s}^4m_T^2 \sinh^2{\eta_s},\nn\\
\delta f^{(7)}&=&0,\nn\\
\delta f^{(8)}&=&0.\nn\\
 \eea

\section{$\delta f$ for Bjorken expansion along with transverse expansion}
\label{App:TransFlow}


Here we calculate the $\delta f$ correction for the case of transverse expansion as 
discussed in section \ref{subsec:TransExpns}. 
Assuming boost invariance, the spatial components of viscous tensor $V^{kl}$ are given by (in order to be consistent with our definition of $V^{kl}$ we neglect the term which  makes $\langle \nabla^{\mu}u^{\nu}\rangle $ traceless  in \cite{Teaney:2003kp}),
\bea
V^{rr}&=&-2\partial_ru^r-2u^rDu^r,\nonumber\\
V^{\phi\phi}&=&-2\partial_{\phi}u^{\phi}-2\frac{u^r}{r}-2r^2u^{\phi}Du^{\phi},\nonumber\\
V^{\eta\eta}&=&-2\frac{u^{\tau}}{\tau},\nonumber\\
V^{r\phi}&=&-r\partial_ru^{\phi}-\frac{1}{r}\partial_{\phi}u^r-ru^rDu^{\phi}-ru^{\phi}Du^{r},\nonumber\\
V^{r\eta}&=&V^{\phi\eta}=0\nonumber,\\
\eea

and the derivatives in the rest frame $Du^{\mu}$ are given by,
\bea
Du^{r}&=&u^{\tau}\partial_{\tau}u^{r}+u^{r}\partial_{r}u^{r}+u^{\phi}\partial_{\phi}u^{r}-r(u^{\phi})^2,\nonumber\\
rDu^{\phi}&=&u^{\tau}\partial_{\tau}(ru^{\phi})+u^{r}\partial_{r}(ru^{\phi})+u^{\phi}\partial_{\phi}(ru^{\phi})+u^{\phi}u^{r}.\nonumber\\
\eea

To fix the values of the time derivatives ($\partial_{\tau}u^{\phi},\partial_{\tau}u^{r},\partial_{\tau}u^{\tau}$) appear in the above equations, it is sufficient to consider the ideal equation of motion. Thus the time derivatives can be determined as following;
\bea
\partial_{\tau}u^{\phi}&=&0\nonumber\\
\partial_{\tau}u^{r}&=&\frac{c_s^2v}{1-c_s^2v^2}\left(\frac{u^{\tau}}{\tau}+\frac{u^{r}}{r}+(1+v^2) \partial_{r}u^r\right)-v\partial_{r}u^r\nonumber\\
\partial_{\tau}u^{\tau}&=&v\partial_{\tau}u^{r}\nonumber\\
\eea

Here $v=u^{r}/u^{\tau}$ is the radial velocity and $c_s$ denotes the velocity of sound.  
We use $c_{s}=0.1$ which is taken  from lattice QCD results at temperature $T_{f}=130$ MeV \cite{Bazavov:2014pvz}.

The corresponding $\delta f^{(i)}$
for the magnetic field $(0,b_{r},b_{\phi},0)$
turns out to be of the following form
 \bea
 \delta f^{(1)}&=& c^{(1)} \tau_{c}\left(\frac{f_0}{T\epsilon}\right)2~
 [p_{r}^{2}V_{rr}+2p_{r}p_{\phi}V_{r\phi}+p_{\phi}^{2}V_{\phi\phi}+p_{\eta}^{2}V_{\eta\eta}],\nonumber\\
 \delta f^{(2)}&=& c^{(2)} \tau_{c}\left(\frac{f_0}{T\epsilon}\right)[(p_{r}^{2}+p_{\phi}^{2}+p_{\eta}^{2})
 (V_{rr}+V_{\phi\phi}+V_{\eta\eta})],\nonumber\\
\delta f^{(3)}&=& c^{(3)} \tau_{c}\left(\frac{f_0}{T\epsilon}\right)
[p_{r}^{2}b_{r}^2V_{rr}+p_{\phi}^{2}b_{\phi}^2V_{\phi\phi}+(p_{r}^{2}+p_{\phi}^{2})b_r b_{\phi}V_{r\phi}\nonumber\\
&+&2p_{r}p_{\phi}b_rb_{\phi}(V_{rr}+V_{\phi\phi})+p_{r}p_{\phi}(b_{r}^{2}+b_{\phi}^{2})V_{r\phi}],\nonumber\\
\delta f^{(4)}&=& c^{(4)} \tau_{c}\left(\frac{f_0}{T\epsilon}\right)[(p_{r}^{2}+p_{\phi}^{2}+p_{\eta}^{2})
 (b_{r}^{2}V_{rr}+b_{\phi}^{2}V_{\phi\phi},\nonumber\\
 &&+2b_{r}b_{\phi}V_{r\phi})],\nonumber\\
\delta f^{(5)}&=& c^{(5)} \tau_{c}\left(\frac{f_0}{T\epsilon}\right)[(p_{r}^{2}b_{r}^{2}+p_{\phi}^{2}b_{\phi}^{2}+2p_{r}p_{\phi}b_rb_{\phi})\nonumber\\
 &&(V_{rr}+V_{\phi\phi}+V_{\eta\eta})],\nonumber\\
\delta f^{(6)}&=& c^{(6)} \tau_{c}\left(\frac{f_0}{T\epsilon}\right)[(p_{r}^{2}b_{r}^{2}+p_{\phi}^{2}b_{\phi}^{2}+2p_{r}p_{\phi}b_rb_{\phi}),\nonumber\\
 &&(b_{r}^{2}V_{rr}+b_{\phi}^{2}V_{\phi\phi}+2b_{r}b_{\phi}V_{r\phi})],\nonumber\\ 
\delta f^{(7)}&=& c^{(7)} \tau_{c}\left(\frac{f_0}{T\epsilon}\right)4p_{\eta}[p_{r}b_{\phi}(V_{rr}-V_{\eta\eta})\nonumber\\
&+&(p_{\phi}b_{\phi}-p_{r}b_{r})V_{r\phi}+p_{\phi}b_{r}(V_{\eta\eta}-V_{\phi\phi})],\nonumber\\ 
\delta f^{(8)}&=& c^{(8)} \tau_{c}\left(\frac{f_0}{T\epsilon}\right)4p_{\eta}[(p_{r}b_{r}^2b_{\phi}+p_{\phi}b_{r}b_{\phi}^2)(V_{rr}-V_{\phi\phi})\nonumber\\
&+&(p_{r}b_{r}b_{\phi}^2-p_{\phi}b_{r}^2b_{\phi}+p_{\phi}b_{\phi}^3+p_{r}b_{r}^3)V_{r\phi}].\nonumber\\
 \eea


\end{document}